\documentclass[3p]{elsarticle}

\usepackage{subfig,multirow,amsmath}

\begin{document}

\begin{frontmatter}
\title{Modeling Financial Volatility in the Presence of Abrupt Changes}
 
 \author[rvt]{Gordon J. Ross\corref{cor1}}
 \ead{gordon.ross@bristol.ac.uk}

 \cortext[cor1]{Corresponding author}
 
 \address[rvt]{Heilbronn Institute for Mathematical Research, University of Bristol, United Kingdom}

\begin{abstract}The volatility of financial instruments is rarely constant, and usually varies over time. This creates a phenomenon called volatility clustering, where large price movements on one day are followed by similarly large movements on successive days, creating temporal clusters. The GARCH model, which treats volatility as a drift process, is commonly used to capture this behavior. However  research suggests that volatility is often better described by a structural break model, where the volatility undergoes abrupt jumps in addition to drift. Most efforts to integrate these jumps into the GARCH methodology have resulted in models which are either very computationally demanding, or which make problematic assumptions about the distribution of the instruments, often assuming that they are Gaussian. We present a new approach which uses ideas from nonparametric statistics to identify structural break points without making such distributional assumptions, and then models drift separately within each identified regime. Using our method, we investigate the volatility of several major stock indexes, and find that our approach can potentially give an improved fit compared to more commonly used techniques.
\end{abstract}

\end{frontmatter}
\section{Introduction}
\begin{figure}
  \centering
 \includegraphics[width=0.8\textwidth]{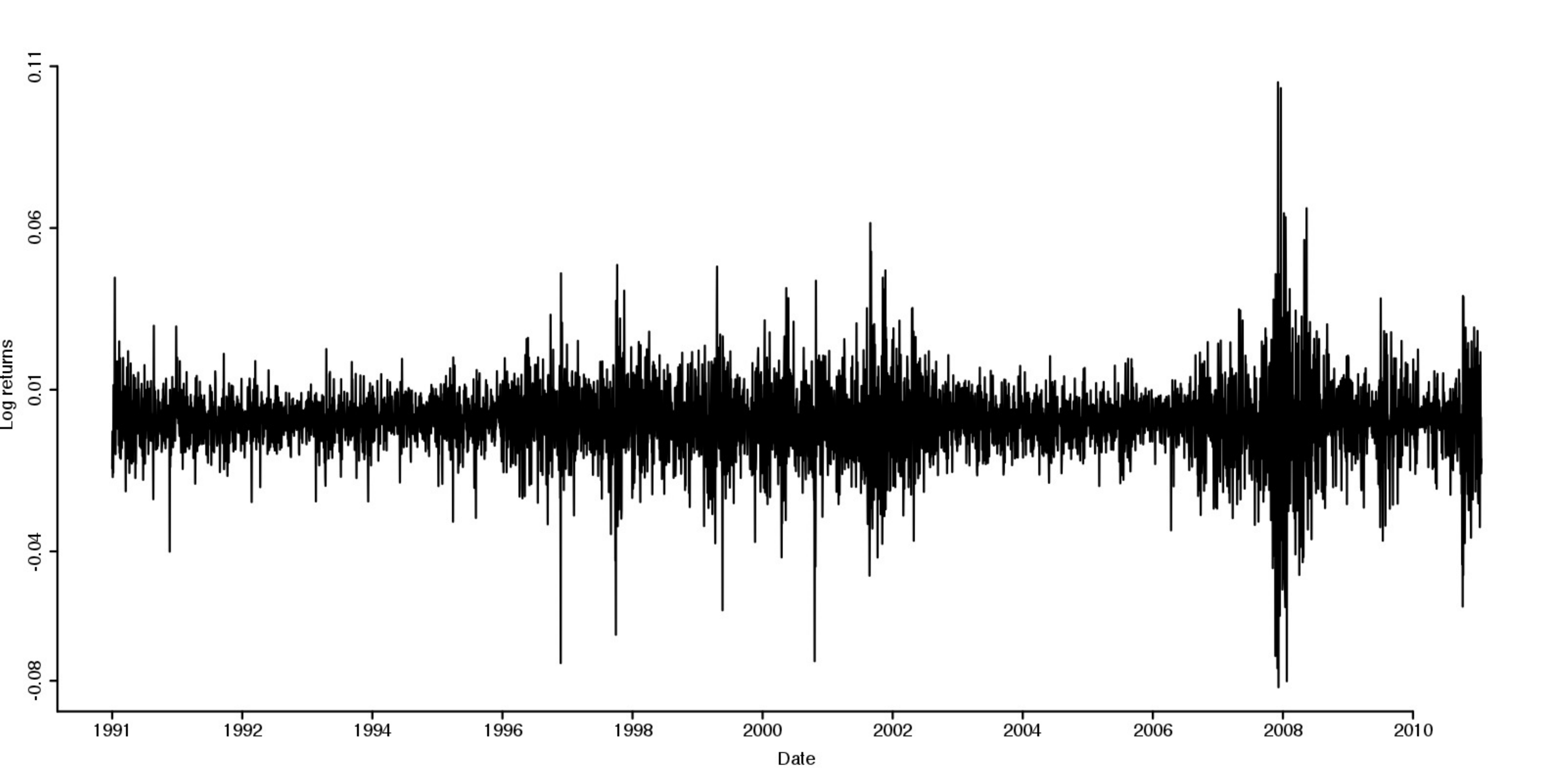}     
\caption{Daily returns of the Dow Jones index between January 1984, and August 2011}
  \label{fig:introftse}
\end{figure}

Volatility clustering is often   observed in the return series of financial instruments \cite{Oh2008,Tsenga2011}. This phenomena is best illustrated by an example. Let $S_t$ denote the price of some financial instrument at a set of equally spaced discrete time points $t = \{1,2,\ldots\}$, and let the \textbf{return series} be the log-increments $r_t = \log S_t - \log S_{t-1}$. The volatility of the instrument is defined as the standard deviation of these returns. A typical example of a financial return series can be seen in Figure \ref{fig:introftse}, which shows the daily returns of the Dow Jones stock index over a $20$ year period ranging from January 1991 to August 2011. It can be observed that the standard deviation is not constant, but instead varies over time. In particular, note that the period from 2003 to 2007 seems to have noticeably lower volatility than the period immediately before or after. Similarly, in $2008$ there are many extreme return values which occur in close succession, pointing to an abnormally high volatility during this period.

\emph{Volatility clustering} refers to this notion that large/small returns tend to be followed by similarly large/small values, which results in extended regimes of abnormally high or low volatility. This has been empirically observed in many different financial time series, and poses a problem for traditional financial models, which have typically assumed that the volatility is roughly constant over time. The last 25 years have seen an increasing number of attempts to model the time-varying nature of volatility, and the generalized  autoregressive conditional heteroskedasticity (GARCH) model \cite{Bollerslev1986}, along with its many variants, is now the de-facto standard. The idea behind GARCH is that the volatility undergoes a stochastic drift process, where the conditional volatility at time $t$ is a random variable, with a conditional distribution  which depends on the long term volatility, the volatility during the most recent period, and the most recent values of the return series.

However the gradual drift process underlying the GARCH model seems to be empirically violated in many real financial series. In some cases, volatility seems to behave more like a jump process, where it fluctuates around some value for an extended period of time, before undergoing an abrupt change, after which it fluctuates around a new value. This can be seen in Figure \ref{fig:introftse} around the year $1996$, where the volatility spontaneously increases for a period of several years, before dropping to a lower value during $2003$. Since the standard GARCH model does not contain the possibility of these sudden jumps, it tends to overestimate the degree of long term volatility persistence. This has prompted the development of regime-switching GARCH processes which can incorporate jumps \cite{Gray1996,He2010}. In these models, the return series  is allowed to contain multiple change points which segments it into  regimes, with the GARCH model having different parameters within each segment. 

However, such models can be hard to estimate. Although there are computationally efficient procedures for estimating multiple change points in simpler ARCH models \cite{Hamilton1994, Kokoszka2000, Lee2004}, the long-range dependence introduced by the GARCH formulation makes such approaches difficult to apply. Standard techniques for fitting multiple change point models to data assume independence between segments \cite{Fearnhead2006} which is not the case in the GARCH framework. Although some recent attempts to fit such models have been attempted \cite{He2010}, it remains a difficult numerical procedure. Therefore, the most popular strategy is to instead use the approximate procedure introduced by  \cite{Aggarwal1999} where the model is fitted in stages, with the abrupt change points first being located using the iterated cumulative sum of squares (ICSS) algorithm \cite{Inclan1994}, before a GARCH model then estimated conditional on these change points. This ICSS-GARCH algorithm has been used  to study a wide variety of financial time series. For example, \cite{Covarrubias2006} uses it to study the volatility of the US dollar exchange rate against several different currencies, \cite{Malik2005} studies the returns of the Canadian stock exchange, \cite{Kang2009} does likewise for the Japanese and Korean exchanges, and \cite{Kang2011} analyses the market for crude oil

Although ICSS-GARCH is simple to implement and has been shown to give improved results compared to standard GARCH models, it is not without its problems \cite{Sanso2004,Rapach2008}. The parameters of the ICSS algorithm are usually designed under the assumption that the financial returns follow a Gaussian distribution, and it can produce many spurious jump points if this assumption is violated. We will show later that applying ICSS to heavy-tailed series can give poor results, since extreme observations are misinterpreted as being regime shifts. Unfortunately, it has now been conclusively established that financial data is very rarely Gaussian, and return series typically exhibit heavy tail behaviour  \cite{Stanley2008,Liu1999,Podobnik2009, Plerou1999, Gopikrishnan1999}. Similar heavy tail behaviour has been observed in many financial series and is not limited to asset returns \cite{Podobnik2011}.

This limitation of the ICSS-GARCH methodology has meant that it is usually only used to detect change points in the weekly returns of financial instruments, i.e. where $S_t$ and $S_{t+1}$ are one week apart \cite{Fernandez2007,Kang2009,Kang2011,Malik2005,Aggarwal1999}. Using the algorithm on daily returns can generate too many spurious false positives for it to be useful, due to the number of extreme values. This is a problem since the daily returns are more fine-grained and hence using them should allow more accurate volatility modelling. Therefore, it is desirable to find a way to use this data when it is available. 

In this paper we present an alternative to the ICSS-GARCH algorithm which is better suited for dealing with the heavy tailed, non-Gaussian data which is typical in finance. We replace the ICSS segmentation step of ICSS-GARCH with an alternative technique based on non-parametric statistics, which does not make any assumptions about the true returns distribution.This allows it to entirely avoid the Gaussianity assumption and allows it to be deployed on daily returns. Our approach is based on the nonparametric change point model framework described in \cite{Rosstechnometrics, Rossjqt}, and we hence refer to it as NPCPM-GARCH. Using this technique, we analyse several stock indexes for volatility change points, specifically focusing on the Dow Jones Industrial Average, the German DAX, the VIX volatility index, and the Japanese Nikkei 225. We compare our results with those of ICSS-GARCH, and find that our method generally gives a better fit to the data when measured using standard criteria. This suggests that it could be a widely useful tool for modelling volatility in other contexts.

The remainder of the paper proceeds as follows. We begin in Section \ref{sec:icssgarch} by describing the ICSS step of the ICSS-GARCH algorithm. We explain why it gives poor performance when used with heavy tailed data, and give a simulated example using Student-t to show this. In Section \ref{sec:cpm} we introduce our new nonparametric approach. We then briefly review the GARCH stage of the algorithm in Section \ref{sec:garch}, and in Section \ref{sec:experiments} we present an empirical evaluation of our method on a range of foreign exchange series.

\section{The ICSS-GARCH Algorithm}
\label{sec:icssgarch}

The ICSS-GARCH methodology  has two stages. Given a financial returns series, the ICSS algorithm is first used to detect any change points in the volatility, and the series is then segmented around these points. Then, a separate GARCH model is fitted to each segment. We begin by giving overview of the ICSS stage, before pointing out its problems and introducing our alternative. Next, we review the GARCH estimation stage.

\subsection{Stage 1: Change Point Detection using ICSS}

The Iterated Cumulative Sum-of-Squares algorithm is based on the work of \cite{Inclan1994}, who proposed a retrospective technique for detecting changes in the variance of a financial time series. Given a series of financial returns $r_1,r_2,\ldots,r_n$ with mean $0$, define the cumulative sum of squares as $C_t = \sum_{i=1}^{n} r_t^2$, and let

$$D_t = \frac{C_t}{C_n} - \frac{t}{n},\quad t = 1,\ldots,n, \quad D_0 = D_n = 0.$$

If the sequence has constant variance, then the value of $D_t$ will oscillate around 0. However if the variance undergoes an abrupt change at some point $\tau < n$ then the value of $D_n$ will exhibit extreme behaviour around this point with its magnitude becoming unusually large. Change detection is carried out by defining a threshold $h_n$, and comparing the maximum value of $D_n$ to this. Specifically, a change is flagged if:

\begin{equation}
\sqrt{n/2}\max_t|D_t| > h_n,
\label{eqn:icss}
\end{equation}
 where the $\sqrt{n/2}$ factor is included for standardization purposes. If the threshold is exceeded then the estimate of the change point, which we denote $\hat{\tau}$, is located at the value of $t$ which gave the maximum value of $|D_t|$, i.e. $\hat{\tau} = \arg\max_t |D_t|$.

In cases where the series may contain multiple change-points, the above procedure can be iterated. The ICSS algorithm is first applied to the full series. If a change is flagged, and estimated to be at location $\hat{\tau}_0$. The series is then split into two segments; $A = \{r_1,r_2,\ldots,r_{\hat{\tau}_0 - 1}\}$ and $B = \{r_{\tau},r_{\tau+1},\ldots,r_n\}$ around this point. Then, the ICSS algorithm is recursively applied to both segments A and B separately, in the same manner as before. If a change point is flagged in segment A, and estimated to be at location $\hat{\tau}_1 < \hat{\tau}_0$, then segment A is further subdivided into two segments around point $\hat{\tau_1}$, and the ICSS algorithm is applied to these new segments. and so on. The same procedure is likewise applied to segment B. This produces a sequence $\hat{\tau}_0,\hat{\tau}_1,\ldots$ of estimated change points.


Deploying the ICSS algorithm requires specifying the threshold $h_n$. In the original paper \cite{Inclan1994}, this is chosen in order to control the probability of mistakenly concluding the a change has occurred, if it fact there is no change. Let $\alpha$ be the probability of this occurring. The authors show that if the observations are Gaussian,then choosing $h_n = 1.358$ asymptotically gives a value of $\alpha=0.05$ assuming that the observations are Gaussian. This is the value which has typically been by other papers using the ICSS-GARCH algorithm \cite{Aggarwal1999,Kang2009,Kang2011}. Note that if the observations are not Gaussian, then the actual value of $\alpha$ obtained for this choice of $h_n$ may be radically different from $0.05$ - this is the crux of the problem with the ICSS algorithm in the context of financial data.






\subsubsection{Non-Gaussian Data}
\label{sec:nongaussian}
The ICSS algorithm is very easy to implement and does not require much computational resources, which is one of the reasons why it has been widely adopted. However its reliance on the Gaussian distribution when specifying the threshold $h_n$ is problematic, since financial data is known to be non-Gaussian, and can exhibit heavy tailed behavior. The justification for the Gaussian assumption comes from the central limit theorem; if there are no change points, then $C_k$ is asymptotically Gaussian since it is a sum of independent and identical random variables. However asymptotic arguments often fail in practice, where we are concerned with finite length return series. The basic problem is that, because financial return series are heavy tailed, there will occasionally be large values generated which are interpreted as change points, even though they should more correctly be classed as outliers.

\begin{figure}
  \centering
  \subfloat[Change points detected by ICSS]{\label{fig:studenticss}\includegraphics[width=0.5\textwidth]{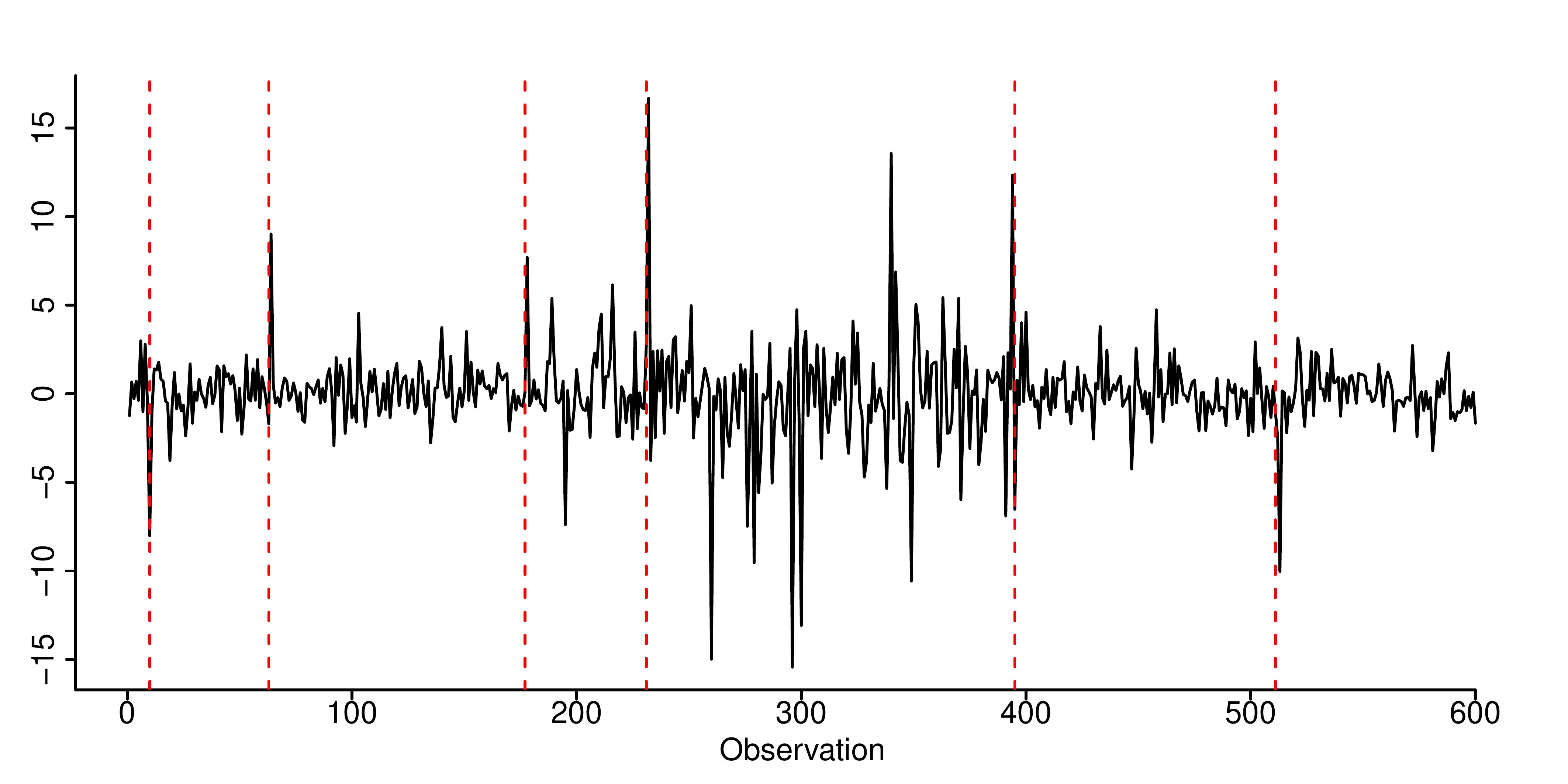}}\\ 
  \subfloat[Change points detected by NPCPM]{\label{fig:studentmood}\includegraphics[width=0.5\textwidth]{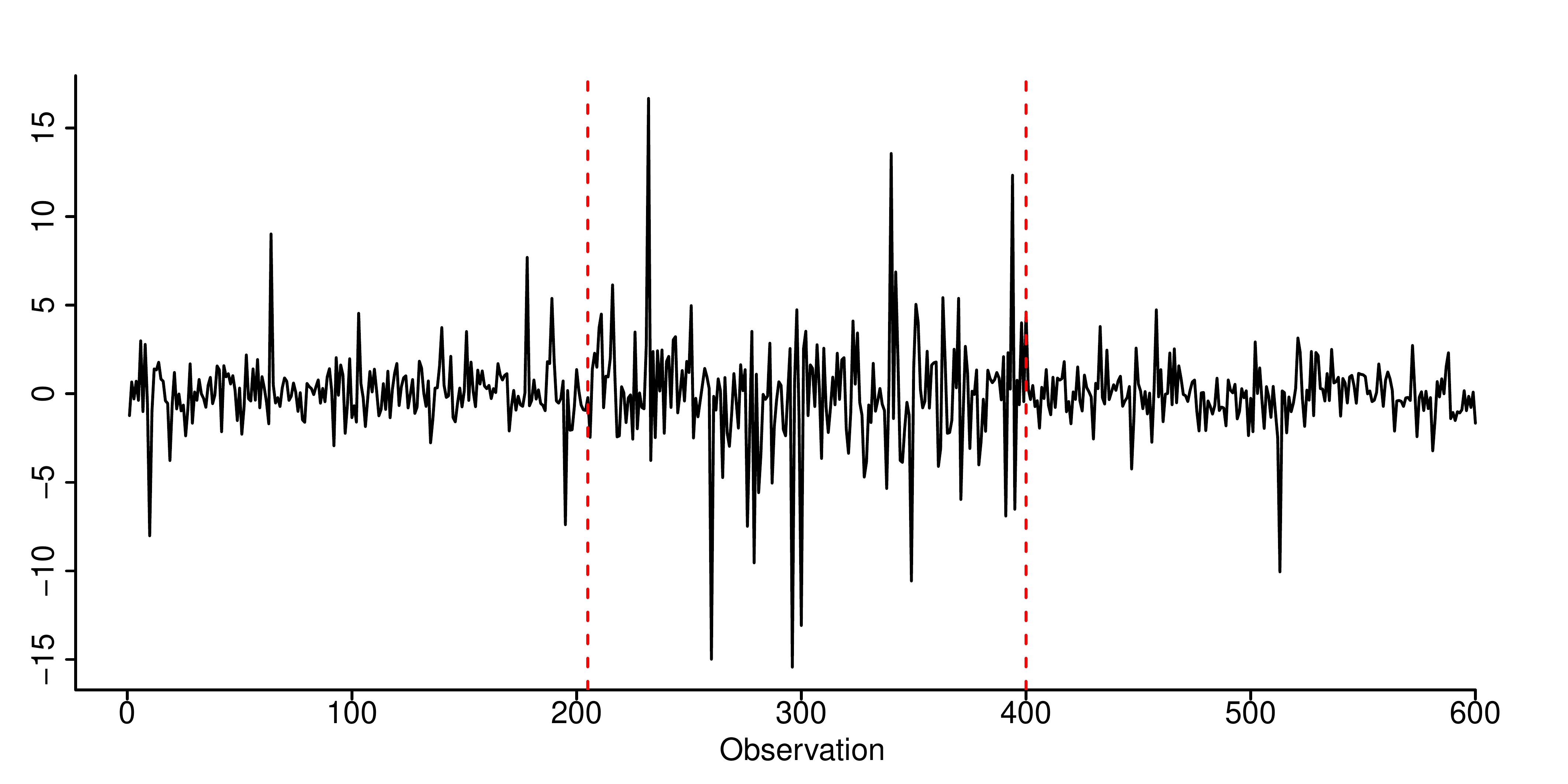}}                  
\caption{The red lines denote the volatility changes identified by the ICSS and NPCPM algorithms in a typical simulated series of Student-t($3$) random variables with true change points at times $200$ and $400$.}
  \label{fig:icssresults}
\end{figure}

We illustrate this by deploying the ICSS algorithm on a simulated series of Student-t random variables, which is a standard distribution used to model heavy tailed behaviour. The series consists of $600$ independent observations . The first $200$ observations have a standard Student-t($3$) distribution, which has mean $0$ and variance of $3$ . The next $200$ observations come from a scaled Student t($3$) distribution with mean $0$ and variance $12$. Finally, the last $200$ observations are again Student-t($3$) with mean $0$ and variance $3$. The series hence consists of $3$ regimes, with a volatility shift in each. We stress that by design the volatility between the change points at times $200$ and $400$ is constant, therefore any change points flagged in these regions are spurious false positives.

We simulated $1000$ realisations of such a series, and applied the ICSS to each. On average, ICSS detected $5.4$ different regimes in each sequence, which is almost three times the true number. A typical realisation of the series is shown in Figure \ref{fig:studenticss}, with the change points discovered by ICSS plotted as red lines, and it can be seen that  many spurious change points are generated. The problem is that the ICSS algorithm is based on the squared magnitudes of the returns, $r_t^2$. However when the observations are heavy-tailed, extreme values can be produced even though nothing in the series has changed. The ICSS algorithm incorrectly interprets these extreme values as being jumps in volatility. This suggests that the ICSS algorithm will not work on daily financial return series which exhibit similar heavy tail behaviour, and our analysis in Section \ref{sec:experiments} will confirm this.

\section{NPCPM: A Nonparametric Alternative to ICSS}
\label{sec:cpm}
The limitation of the ICSS algorithm is hence the assumption that the returns are Gaussian. We therefore propose replacing the ICSS stage of ICSS-GARCH algorithm with a new technique which makes no such distributional assumptions, based on the sequential change detection work in \cite{Rosstechnometrics}. We call this approach the Nonparametric Change Point Model (NPCPM). Recall from the discussion of ICSS that it was configured to have a $0.05$ probability of incurring a false positive, given that there are no changes in the return series, and that the Gaussian assumption holds.  We wish to retain this fixed $0.05$ false positive probability, regardless of the true return distribution. This can be done by adopting the idea of \textbf{rank tests} from the field of nonparametric statistics. These tests are easy to understand and implement, yet are very powerful and able to maintain a fixed rate of false positives regardless of the true distribution of the return series. We first review a standard non-parametric test for comparing two samples of observations and testing whether they have the same variance, when their distribution is unknown. Then, we show how this test can be extended to detect changes in volatility.
	
\subsection{Two Sample Testing}
Suppose that we have two samples of observations $A = \{r_{1,1},r_{1,2},\ldots,r_{1,m}\}$, $B=\{r_{2,1},r_{2,2},\ldots,r_{2,n}\}$, with an unknown heavy-tailed distribution, and we wish to test whether they have equal variance. One commonly used method for this is the Mood test \cite{Mood1954} . This consists of replacing each observation with its \textbf{rank}, which is defined as the number of observations in the combined sample which it is greater than. More formally, the rank of each observation $r_i$ is:

$$
\text{rank}(r_i) = \sum_j^m  I(r_i \geq r_{1,j})  +  \sum_j^n I(r_i \geq r_{2,j}), \quad \text{ where } I(r_i \geq r_j) = \left\{ \begin{array}{rl}
 1 &\mbox{ if $r_i \geq r_j$} \\
 0 &\mbox{ otherwise}
       \end{array} \right. 
$$

So for example, if the first sample contains the observations $(1.02, 1.32, 2.17)$ and the second sample contains $(0.87, 1.21, 1.89)$ then the observation $0.87$ has rank $1$, the observation $1.02$ has rank $2$, and so on. The key point in the theory of rank tests is that if both samples have the same distribution, then each observation is equally likely to have any of the $m+n$ possible ranks. This is true regardless of what the true distribution is, and no matter how heavy tailed it is. Therefore, any test statistic which depends only on the ranks of the observations will not depend on their distribution.

The Mood test for equal variance measures the extent to which the rank of each observation deviates from the median rank. If both samples have an identical distribution, then the median rank is simply $(n+m+1)/2$. If the observations all have the same distribution, then we would expect the ranks to be roughly equally split between the two samples. However if the variance of the samples differs, then the one with higher variance will typically  significantly more extreme observations than the other. This leads to a test statistic based on summing the squared rank deviations from either of the samples, and comparing it to a threshold:

$$M_{m,n}' = \sum_{i=1}^m (\text{rank}(r_{1,i}) - (n+m+1)/2)^2.$$

The expected value and standard deviation of this $M'_{m,n}$ statistic depends on the sample sizes $m$ and $n$. To make it easier to compare values evaluated on different sample sizes, we standardise it by subtracting its mean and dividing by its variance. From \cite{Mood1954} this can be shown to be:

$$M = (|M' - \mu_{M'}|)/\sigma_{M'}, \quad \quad  \mu_{M'} = m(N^2 - 1)/12, \quad \sigma^2_{M'}= mn(N+1)(N^2-4)/180, \quad N = m+n.$$

Finally if $M_{m,n}' > h_n$ for some appropriately chosen threshold, then we conclude that the two samples have unequal variance. The value of $h_{m,n}$ can again be chosen to (e.g.) give a $0.05$ probability of falsely concluding that the samples have unequal variance when they are in fact equal. Unlike the ICSS approach, this threshold can be chosen in a way that allows this probability to hold regardless of how the returns are distributed.

\subsection{Change Detection}

Given the return series $r_1,\ldots,r_t$, we wish to test whether there is a change in volatility. Assuming for now that there is at most a single change point, we can think of this as being a compound problem; we first test whether there is a change point immediately after the second observation, then test if there is a change point after the third observation, and so on. More formally, we wish to decide between the hypothesis that there is no change point in the series, and the hypothesis that there is a change point at observation $r_k$ for some unknown value of $1<k<t$.

For each possible value of $k$, split the observations into two samples $\{r_1,r_2,\ldots,r_k\}$, $\{r_{k+1},r_{k+2},\ldots,r_t\}$. Then, the Mood test can be applied to these two samples in order to compare whether they have equal variance as before. Let $M_{k,n}$ be the computed value. By repeating this procedure over all values of $k$, the following maximized test statistic can be defined:

$$M_t = \max_k M_{k,n}$$.

The  test then consists of comparing this maximized statistic to a threshold $h_t$. As before, if $M_n > h_n$ for an appropriate threshold, we conclude that a change has occurred, with the best estimate of the change point then being $\hat{\tau} = \arg\max_k |D_k|$.  Note the similarity between this, and the ICSS statistic in equation \ref{eqn:icss}. In both cases, we are essentially performing a test at each individual point in the sequence, and  picking out the value which maximises it. 

\begin{table}[t]
\centering
\begin{tabular}{ |c|cccccccccc|}
\hline
n & 10 & 20 & 50 & 100 & 200 & 500 & 1000 & 5000 & 10000 & 20000\\
\hline
$h_n$ & 2.48 & 2.65 & 2.88 & 2.99 & 3.09 &  3.20 & 3.25 & 3.35 & 3.37 & 3.42\\
\hline
\end{tabular}

\caption{Values of the $h_n$ threshold which give a $0.05$ probability of incurring a spurious false positive when using the maximized Mood statistic, for various lengths $n$ of the financial series.}
\label{tab:moodthresholds}
\end{table}

The final step is specifying the value of $h_n$. Similar to the ICSS algorithm, we wish to choose this so that the probability of incurring a false positive is equal to either $0.05$; it should therefore be chosen as the $95^{th}$ percentile of $N_n$. Unlike this ICSS algorithm, doing this will guarantee a false positive probability of $0.05$ regardless of the return distribution. These values can be easily found using Monte Carlo simulation. In Table \ref{tab:moodthresholds} we list the $h_n$ values which give a false positive probability of $0.05$, for various lengths of the financial series.

In cases where the series may contain multiple change-points, we use the same recursive approach as in the ICSS algorithm. We first run our method on the whole series, and compare $M_t$ to the threshold $h_n$. If it exceeds it, let $\hat{\tau_0} = \arg\max_k |D_k|$. The observations are then split into two samples around $\hat{\tau_0}$, and the change detection algorithm is recursively applied to each sample until the threshold is no longer exceeded. 

To illustrate the advantage of our approach over ICSS when working with non-Gaussian data, we applied it to the same heavy-tailed Student-t data discussed in the previous section. Based on $10000$ simulations, the NPCPM algorithm on average detects $2.1$ change points per sequence, compared to both the true number of $2$, and the average of $5.4$ found by the ICSS algorithm. This highlights that the NPCPM approach is much more accurate when working with non-Gaussian data. Figure \ref{fig:studentmood} shows the volatility change points which are identified in a typical realisation of the Student-t($3$) series. Unlike the ICSS algorithm, our approach does not typically generate spurious false positives even though the observations are heavy-tailed. This shows that it is better able to cope with heavy tailed observations, and should be better suited to financial data.

\subsection{Stage 2: GARCH Modelling}
\label{sec:garch}

After the jump points have been found using either ICSS or NPCPM, the next step is to model the volatility drift in the segments between each pair of change points. This is done using the GARCH(\emph{p},\emph{q}) model \cite{Bollerslev1986}, where the conditional variance of the returns obeys an autoregressive moving average process, with $p$ and $q$ denoting the time lags. In practice, by far the most common version of this model is the GARCH(1,1), which we also use. A financial time series is said to be GARCH(1,1) if its volatility $h_t$ has the following time-varying form:
$$r_t \sim h_t \epsilon_t$$
$$h_{t} =  \omega + \alpha h_{t-1} + \beta r_{t-1}^2,$$
where $\epsilon_t$ is a sequence of independent and identically distributed random variables. In other words, the volatility at time $t$ is a function of the long term volatility ($\omega$), the variance at the previous time point ($h_{t-1}$), and the squared previous return ($r_{t-1}$). This reliance on previous values leads naturally to the volatility clustering effect as seen in Figure \ref{fig:introftse}. The distribution of $\epsilon_t$ is often taken to be Gaussian $N(0,1)$, but we will also consider the case where the $\epsilon_t$ variables have a Student-t distribution with $v$ degrees of freedom as in \cite{Bollerslev1987}, in order to model heavy tail behaviour. 


One limiting feature of the GARCH model is that the volatility is mean reverting and fluctuates around a fixed value. As discussed in the Introduction, it is often more realistic to use a regime-switching/change point formulation where the parameters of the GARCH model, and hence the long-run volatility, can take different values in each segment. In this case, the segment boundaries are the change-points found by the ICSS or NPMLE algorithms. We consider two different GARCH models; the first is the one used in \cite{Aggarwal1999, Malik2005, Kang2011} where only the $\omega$ parameter undergoes change, i.e:
$$h_{t} =  \omega_t + \alpha h_{t-1} + \beta r_{t-1}^2,$$
where $\omega_t$ is equal to some constant $k_0$ until the first change point, before switching to $k_1$ until the next change point, and so on. In the second regime-switching model all three parameters are allowed to vary between regimes, i.e:

$$h_{t} =  \omega_t + \alpha_t h_{t-1} + \beta_t r_{t-1}^2.$$

This gives a more flexible model, at the risk of overparameterization. We will refer to the model where only $\omega$ changes as the $\omega$-GARCH model, and the one where all parameters change as $\omega \alpha \beta -$GARCH. Note that when using the models with Student-t error with $v$ degrees of freedom, we treat $v$ as a free parameter which is estimated along with the GARCH coefficients.

\begin{figure}[!t]
  \centering
  \subfloat[Dow Jones]{\label{fig:gbp}\includegraphics[width=0.5\textwidth]{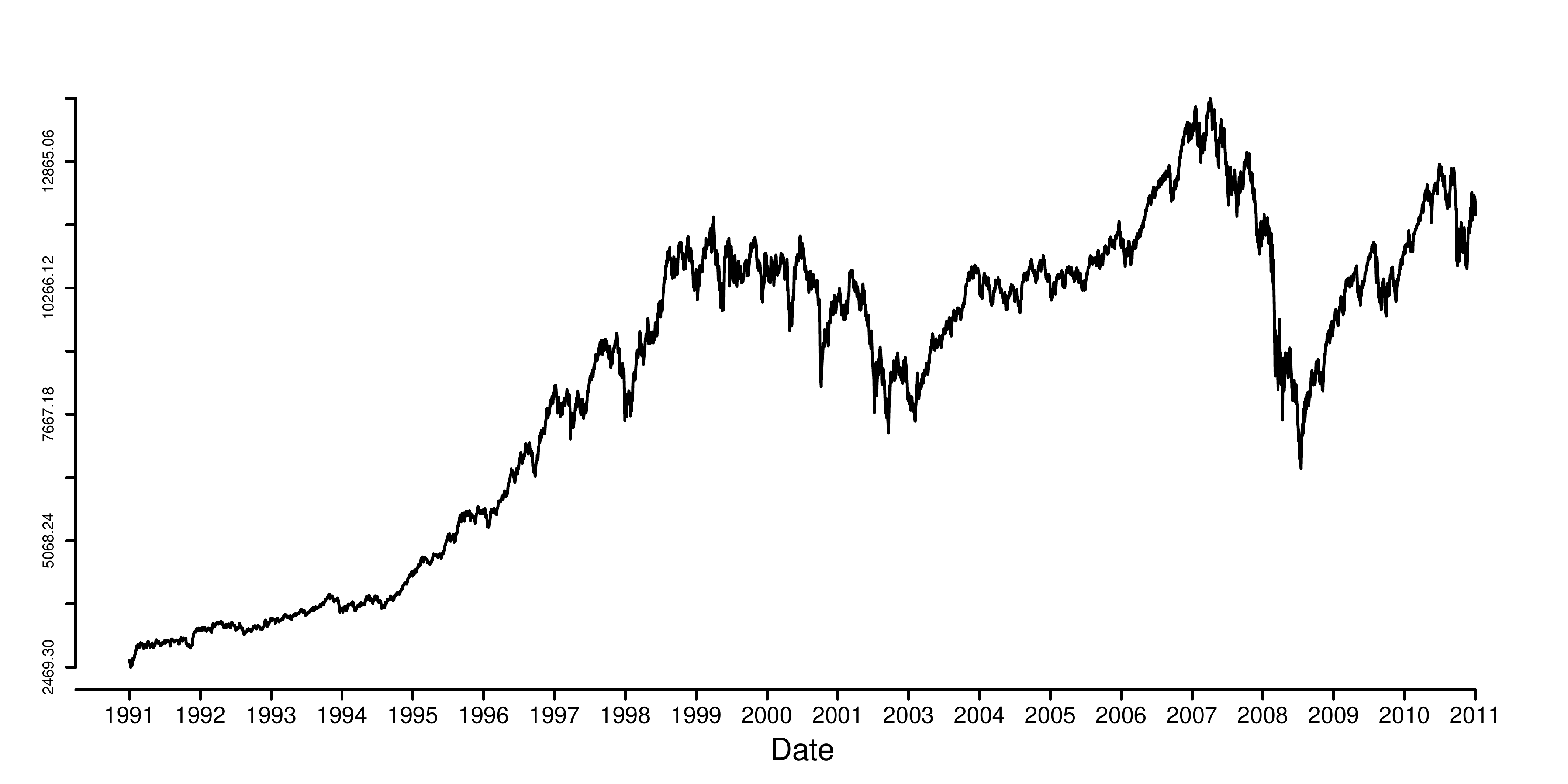}}           
  \subfloat[DAX]{\label{fig:euro}\includegraphics[width=0.5\textwidth]{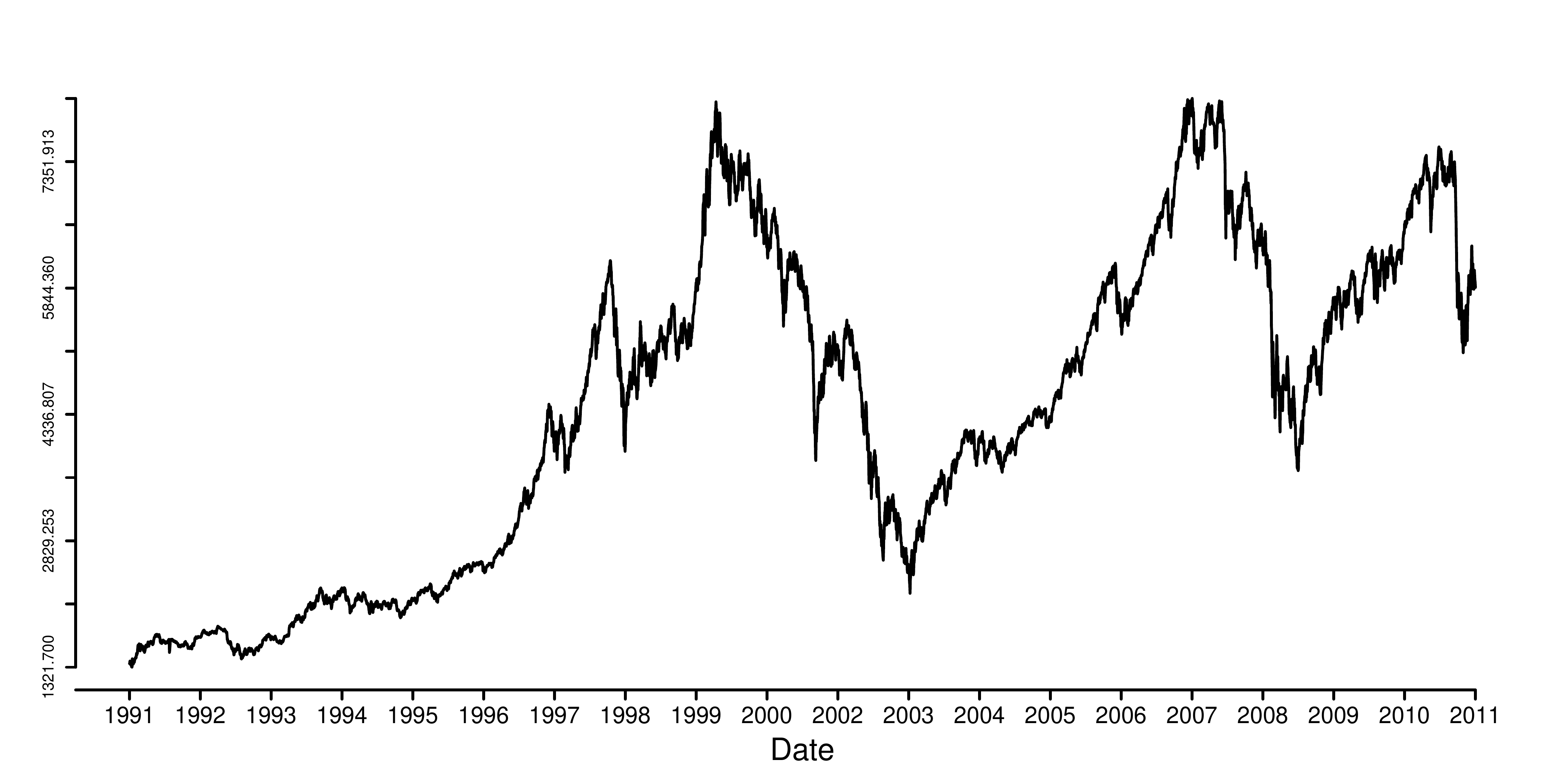}}\\
  \subfloat[Nikkei]{\label{fig:chf}\includegraphics[width=0.5\textwidth]{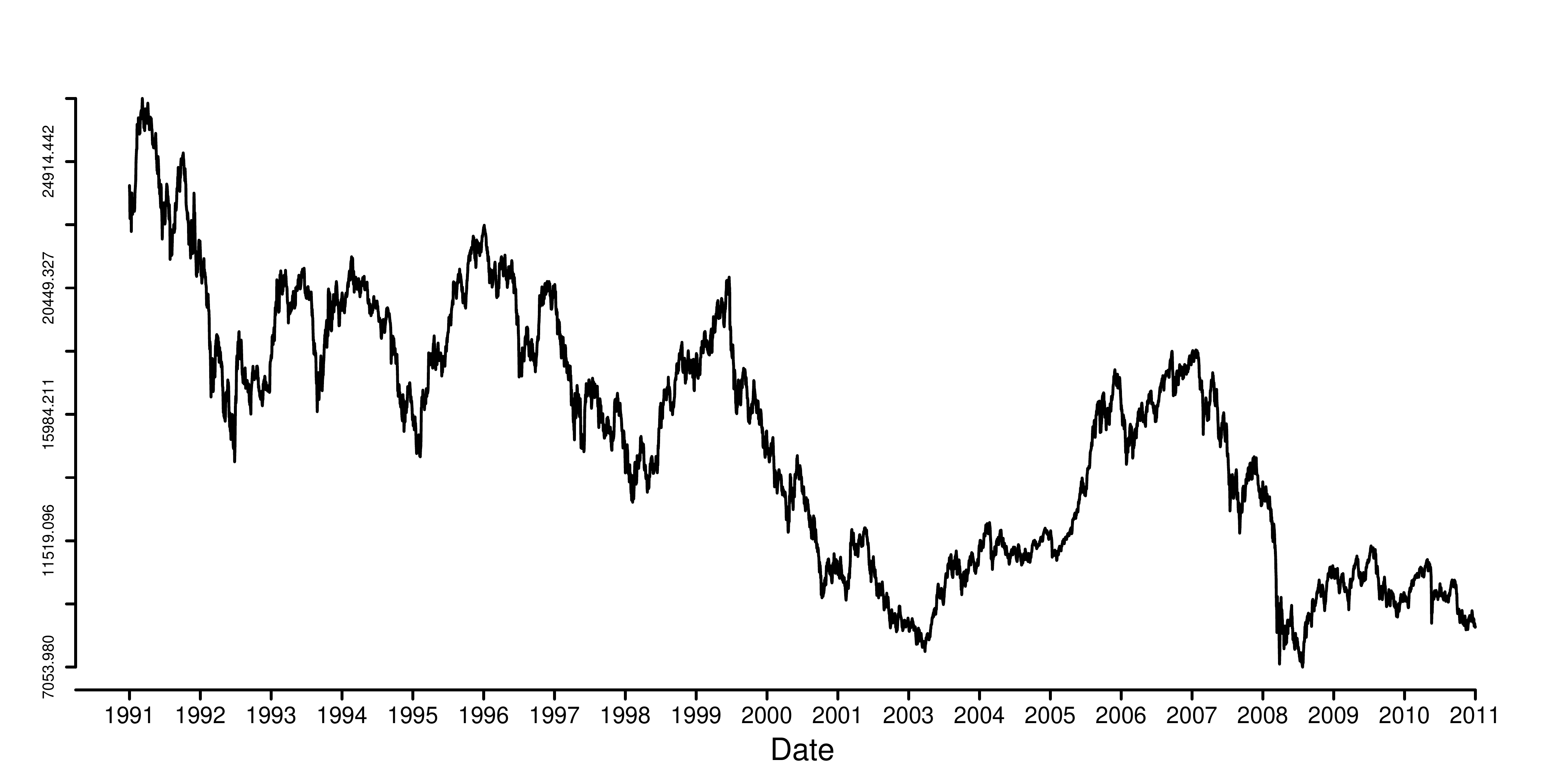}}           
  \subfloat[VIX]{\label{fig:sek}\includegraphics[width=0.5\textwidth]{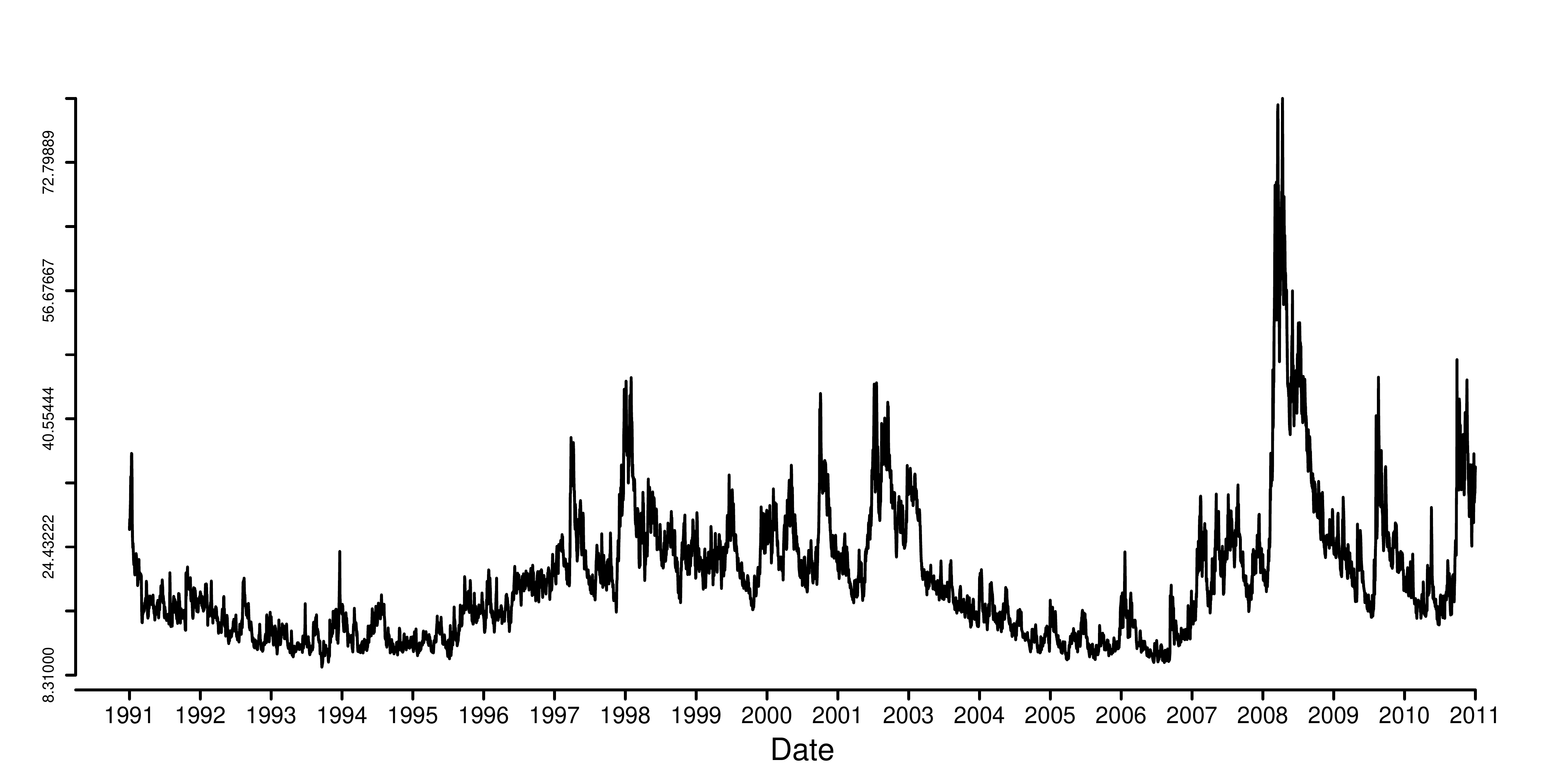}}\\
\caption{Daily closing prices for the stock indexes we are investigating}
  \label{fig:priceseries}
\end{figure}

\section{Empirical Results}
\label{sec:experiments}

Our change point study uses four world stock indexes, which are 1) the Dow Jones Industrial Average which consists of 30 large companies in the United States, 2) the Deutscher Aktien Index (DAX) which consists of the 30 large Germany companies, 3) the Nikkei 225 which consists of 225 Japanese countries, and 4) the VIX volatility index, which measures the implied volatility of the companies from the S$\&$P 500. We obtained daily closing prices for each series between the $1^{st}$ of January 1991, and the $31^{st}$ of October 2011. Figure \ref{fig:priceseries} shows a plot of each of these four series.

\begin{table}[!b]
\centering
\small{
\begin{tabular}{|l|l l l l|} 
\hline
		&Dow Jones		&DAX 			&Nikkei			&VIX\\	
\hline 
Mean		&$-2.87\times 10^{-4}$	&$-2.75 \times 10^{-4}$	&$-2.03 \times 10^{-4}$	&$4.17 \times 10^{-5}$ \\
Standard Dev	&$1.12 \times 10^{-2}$	&$1.45 \times 10^{-2}$	&$1.41 \times 10^{-2}$	&$6.00 \times 10^{-2}$ \\
Skew		&0.15			&0.33			&$0.01$			&-0.44\\
Kurtosis	&8.09			&5.12			&4.16			&$2.48$\\
\hline
Shapiro Wilk	&$4.51 \times 10^{-46}$	&$4.44 \times 10^{-41}$	&$2.83 \times 10^{-33}$	&$4.24 \times 10^{-29}$\\
Ljung Box	&$4.84 \times 10^{-4}$	&0.67			&$1.04 \times 10^{-3}$	&$3.61 \times 10^{-8}$\\
Ljung Box$^2$	&$< 1.00 \times 10^{-16}$ &$< 1.00 \times 10^{-16}$	&$< 1 \times 10^{-16}$	&$2.20 \times 10^{-15}$\\
\hline 
\end{tabular}
}
\caption{Summary statistics for the series of daily differences for the four stock indexes. The bottom half shows the p-values obtained for several statistical tests, where the $^2$ symbol denotes that it has been applied to the squared differences.}
\label{tab:summary}
\end{table}

For each series, we analyze the logarithm of the daily price differences defined as $r_t = \log S_t - \log S_{t-1}$. Table \ref{tab:summary} displays some summary statistics for each sequence of differences. It can be seen that all have a mean of near 0, as should be expected. All series exhibit kurtosis far in excess of what would be expected if they followed a Gaussian distribution (recall that the Gaussian distribution has a kurtosis of $0$).  To test this further, we show the p-values of the standard Shapiro-Wilk test \cite{Shapiro1965} for Gaussianity. The small p-values show that the Gaussian hypothesis should be rejected for all series. Finally, we give the p-values associated with the Ljung Box test for autocorrelation, in both the original series of differences, and their squared values. If the volatility of each series was constant then we would expect there to be no autocorrelation in the squared differences; the low p-values obtained for all series show that this hypothesis should be rejected, and that the volatility is not constant. 

We next investigate the change points which are found by the ICSS and NPCPM algorithms in these series. After this, we will fit the full GARCH model, and compare these two methods more formally.

\subsection{Change Point Analysis}

\begin{figure}
  \centering
  \subfloat[Dow Jones (22 change points)]{\label{fig:gbp05icss}\includegraphics[width=0.5\textwidth]{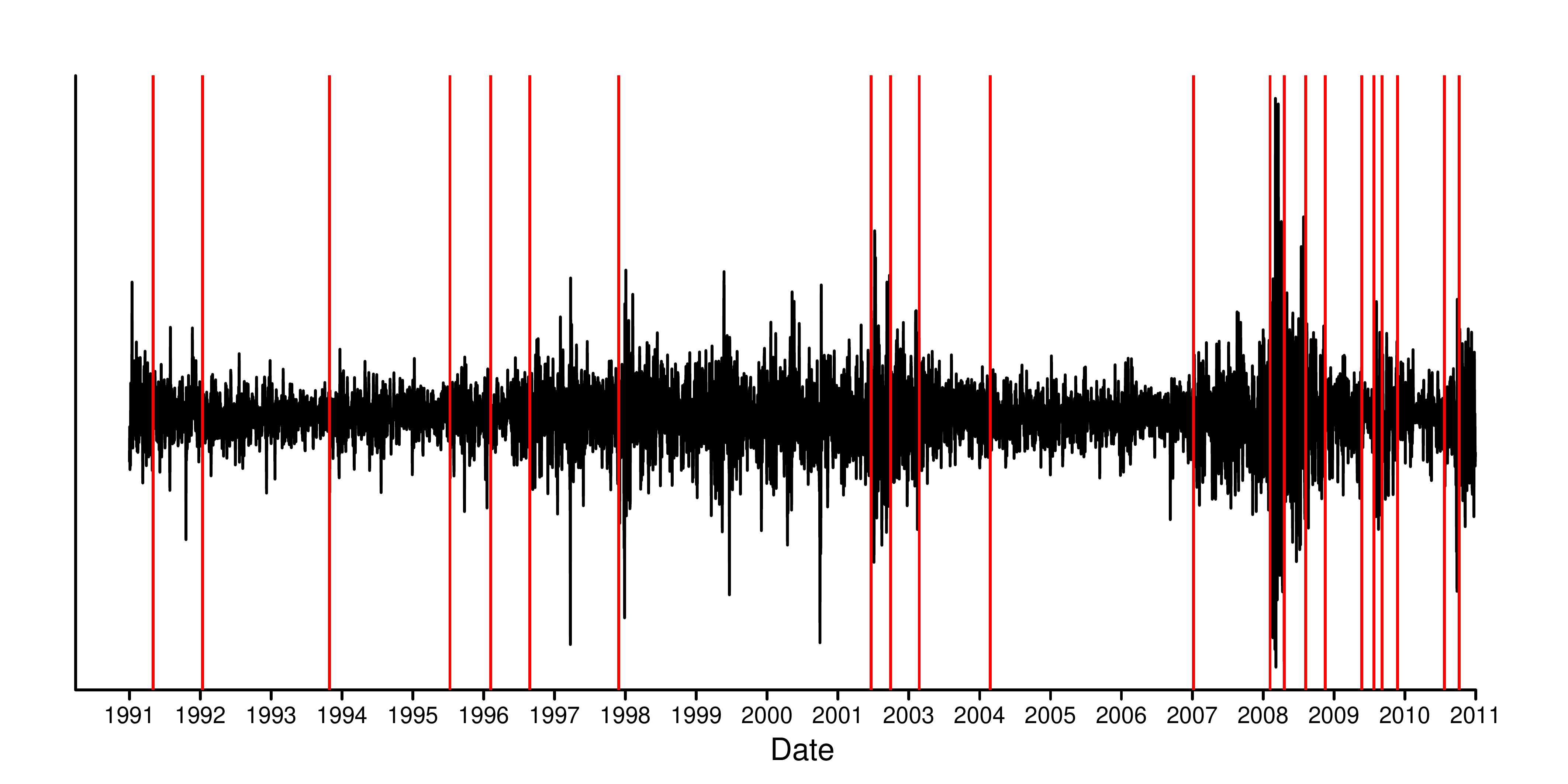}}
  \subfloat[DAX (45 change points)]{\label{fig:euro05icss}\includegraphics[width=0.5\textwidth]{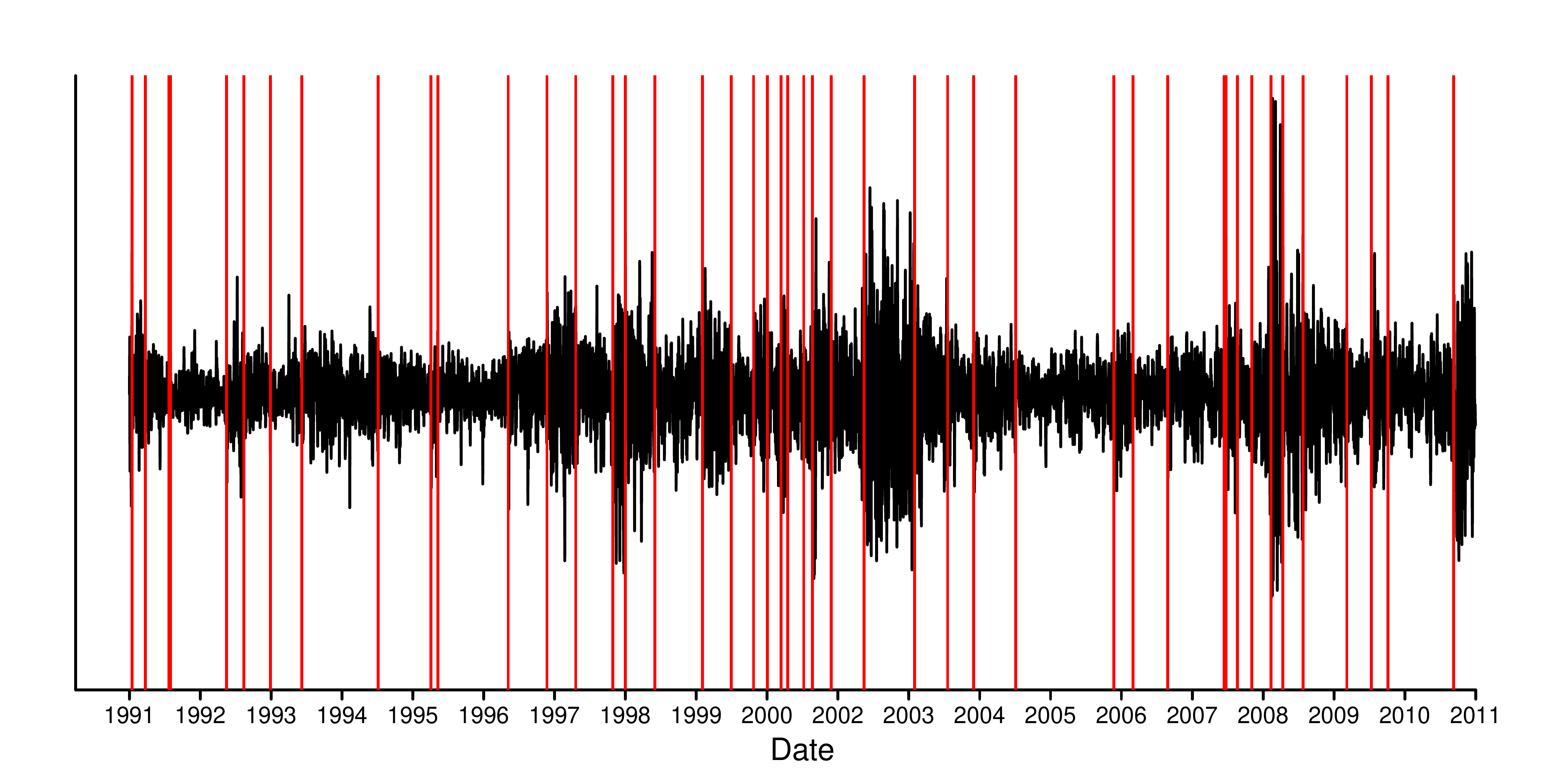}}\\    
  \subfloat[Nikkei (31 change points)]{\label{fig:chf05icss}\includegraphics[width=0.5\textwidth]{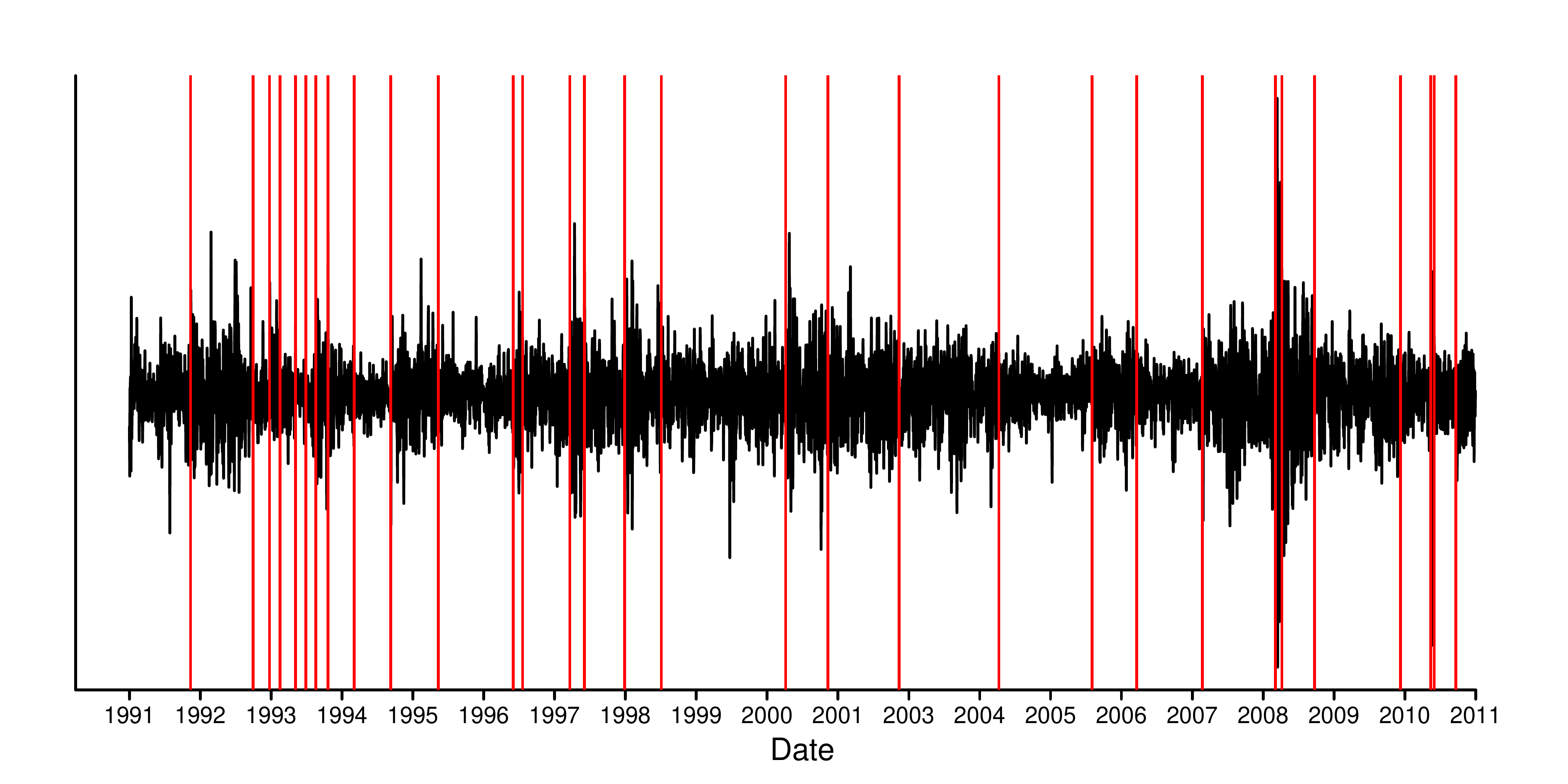}}       
  \subfloat[VIX (30 change points)]{\label{fig:vix05icss}\includegraphics[width=0.5\textwidth]{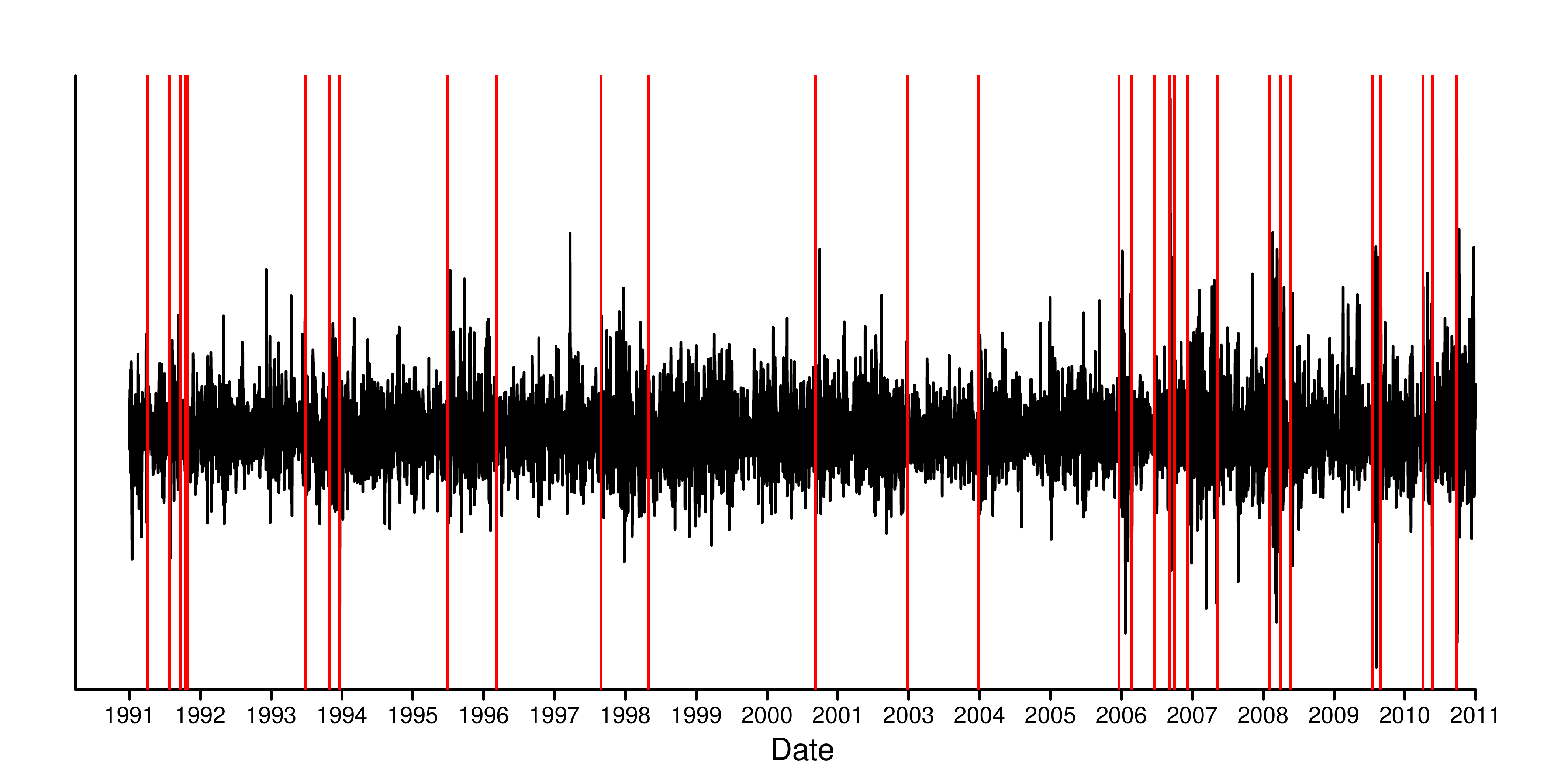}}\\
\caption{Change points discovered in the daily differences for each stock index using the ICSS algorithm}
  \label{fig:changepointsicss}
\end{figure}

We begin by investigating the change points which are discovered by both the ICSS and NPCPM algorithms. We configured both algorithms to have a significance level of $\alpha=0.05$ as discussed previously. In Figure \ref{fig:changepointsicss} we show the change points which were detected by the ICSS algorithm. It can be seen that there are a very large number of change points detected, with $45$ for the Dax index, and 31, 30, and 22 for the others respectively. Most of these do not seem to correspond to genuine long term changes in the volatility; as we would expect from our discussion in Section \ref{sec:nongaussian}, many seem to be false positives flagged in response to the extreme values for the daily differences which sometimes .

\begin{figure}
  \centering
  \subfloat[Dow Jones (12 change points)]{\label{fig:gbp05mood}\includegraphics[width=0.5\textwidth]{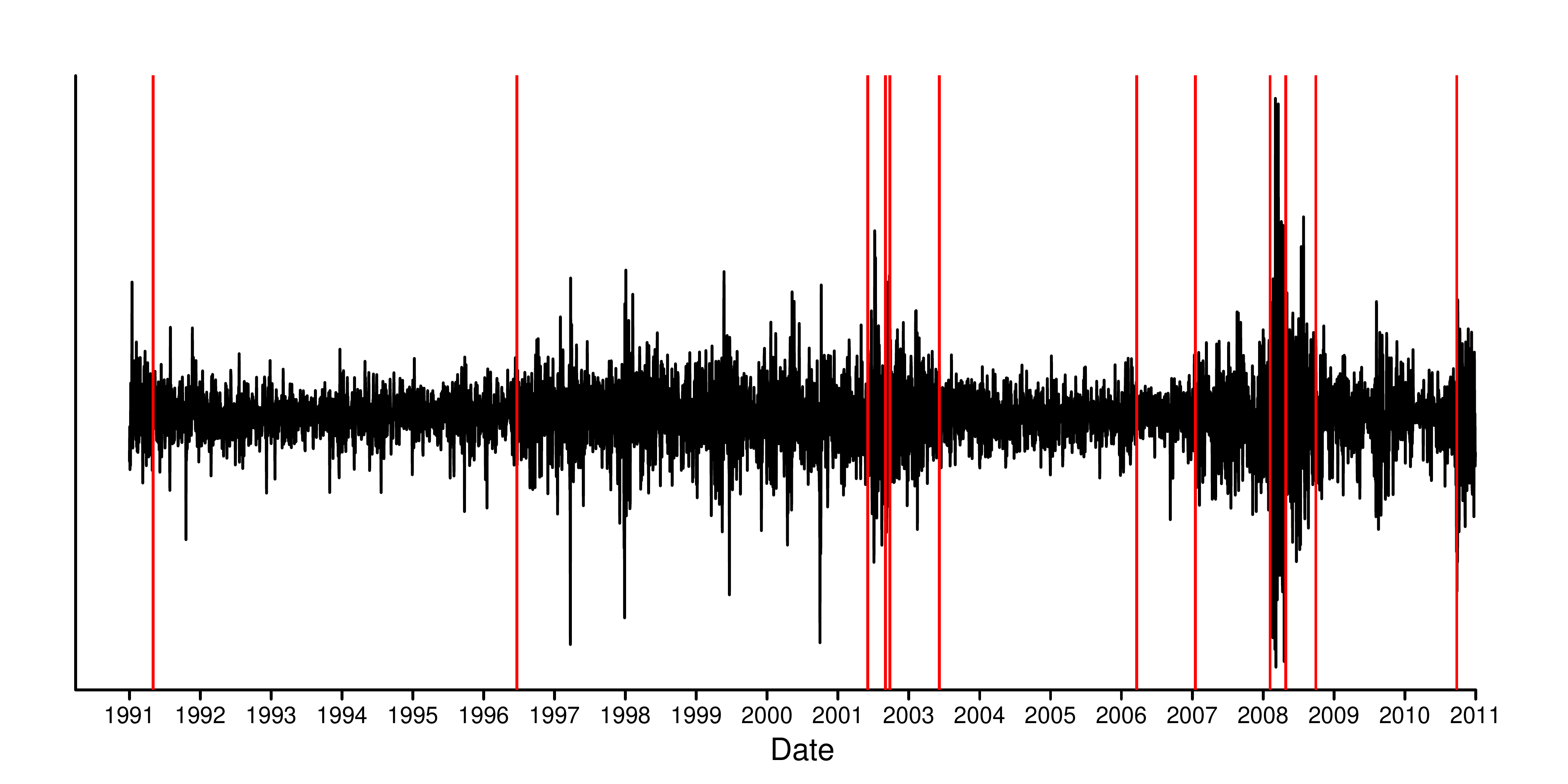}}    
  \subfloat[DAX (21 change points]{\label{fig:euro05mood}\includegraphics[width=0.5\textwidth]{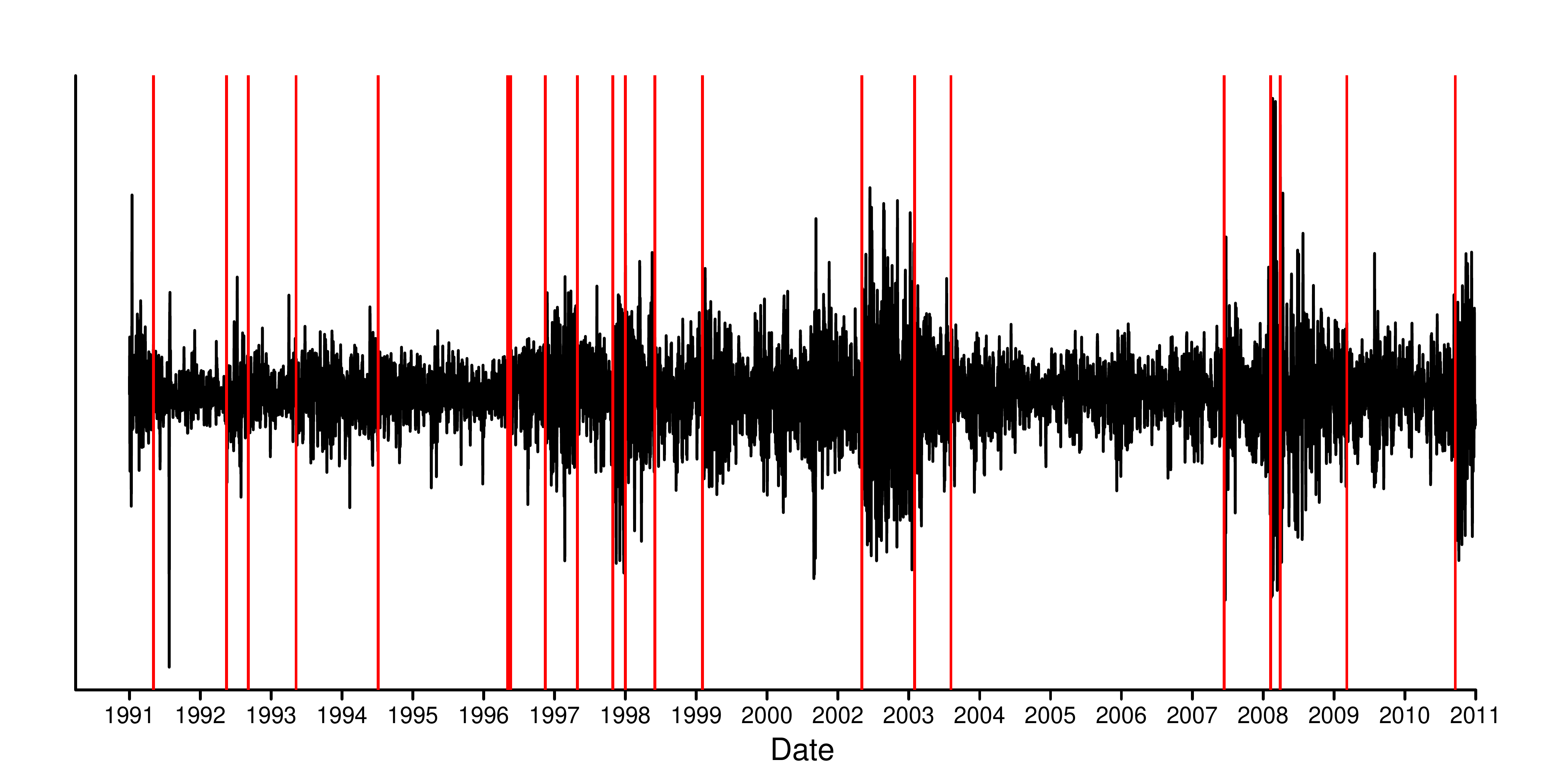}}   \\
  \subfloat[Nikkei (21 change points)]{\label{fig:chf05mood}\includegraphics[width=0.5\textwidth]{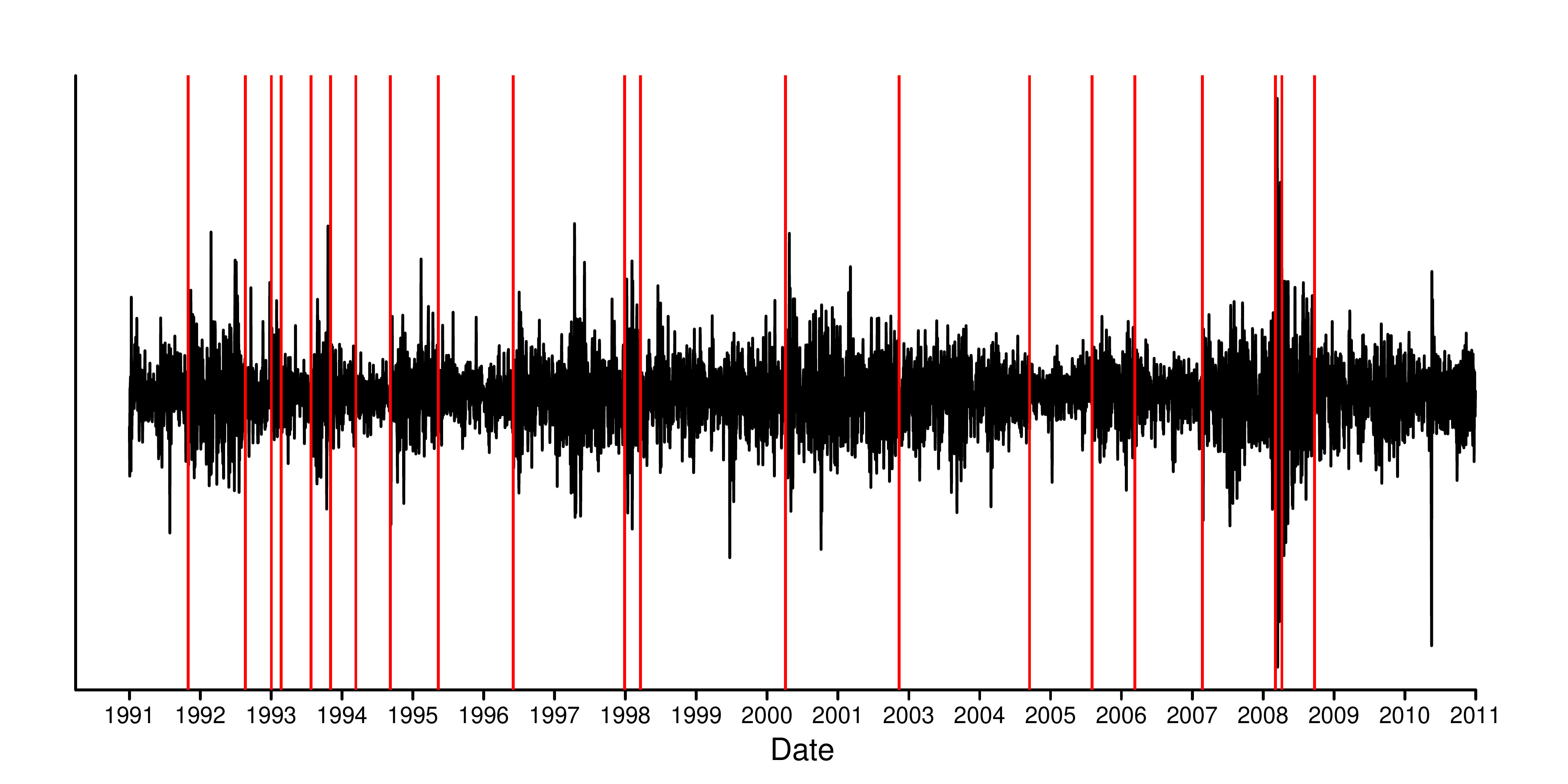}}   
  \subfloat[VIX (10 change points)]{\label{fig:vix05mood}\includegraphics[width=0.5\textwidth]{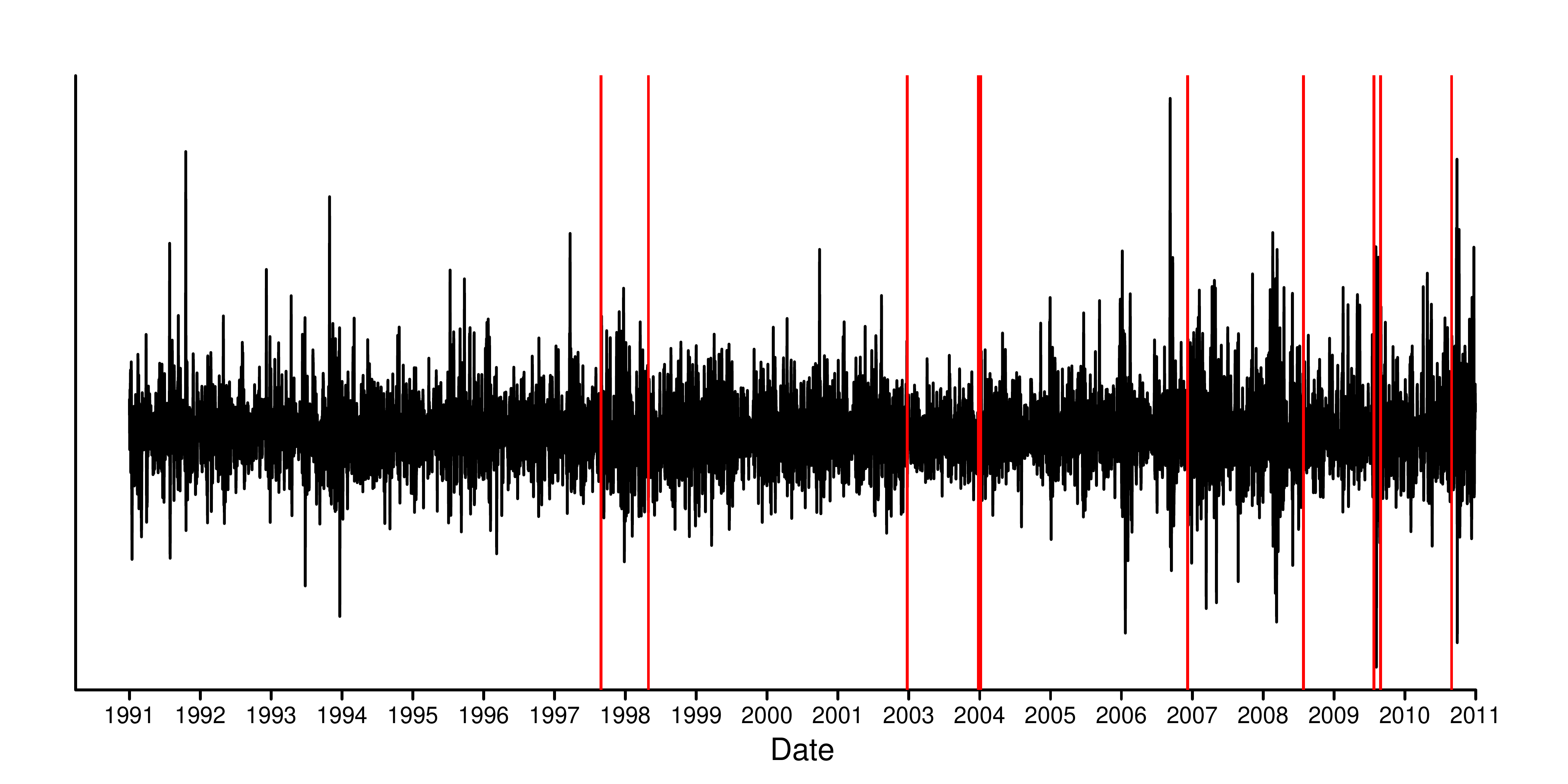}}  
\caption{Change points discovered in the daily differences for each stock index using the NPCPM algorithm}
  \label{fig:changepointsmood}
\end{figure}

In Figure \ref{fig:changepointsmood} we show the results of the NPCPM algorithm applied to the same series. In contrast to the ICSS, there are fewer change points detected, suggesting that this is giving a better fit to the data. Unlike ICSS, the change points found by NPCPM do not seem to correspond to the outlying observations, suggesting robustness. In the following section we will use standard model fitting criteria to give a more quantitative determination of which algorithm is more accurately finding change points.

\begin{table}[!b]
\centering
\tiny{
\begin{tabular}{|c|c c c|c c c|c c c|c c c|} 
\hline
& \multicolumn{3}{|c|}{Dow Jones} &\multicolumn{3}{|c|}{DAX} &\multicolumn{3}{|c|}{Nikkei} &\multicolumn{3}{|c|}{VIX}\\	
 & L & AIC & BIC & L & AIC & BIC & L & AIC & BIC & L & AIC & BIC \\
\hline 
$\omega-$GARCH, ICSS, Gaussian & 16908& -33761& -33584& 15561& -31024& -30702& 7480& -14895& -14685& 13691& -27311& -27082\\
$\omega-$GARCH, ICSS, Student-t & 16336& -32617& -32433& 15057& -30014& -29685& 7002& -13938& -13721& 13845& -27618& -27382\\
$\omega-$GARCH, NPCPM, Gaussian & 16862& -33688& -33570& 15245& -30435& -30251& 7573& -15113& -15008& 13756& -27463& -27306\\
$\omega-$GARCH, NPCPM, Student-t & 16362& -32685& -32560& 14691& -29325& -29134& 7090& -14146& -14034& 13755& -27461& -27297\\
\hline
$\alpha\beta\omega$-GARCH, ICSS, Gaussian & 17229 & -34277 & -33679 & 15958 & -31549 & -30346 & 14855 & -29440 & -28556 & 7761 & -15276 & -14468 \\ 
$\alpha\beta\omega$-GARCH, ICSS, Student-t & 17333 & -34483 & -33885 & 15992 & -31619 & -30416 & 14903 & -29536 & -28653 & 7885 & -15525 & -14717 \\ 
$\alpha\beta\omega$-GARCH, NPCPM, Gaussian & 17178 & -34255 & -33920 & 15836 & -31499 & -30927 & 14830 & -29485 & -28916 & 7630 & -15173 & -14891 \\ 
$\alpha\beta\omega$-GARCH, NPCPM, Student-t &  17307 & \textbf{-34512} & -34177 & 15944 & \textbf{-31714} & -31143 & 14878 & \textbf{-29582} & -29012 & 7823 & \textbf{-15560} & -15278 \\ 
\hline
$\alpha\beta\omega$-GARCH, GICSS, Gaussian & 17162 & -34294 & -34195 & 15800 & -31537 & -31334 & -14693 & -29380 & -29360 & 7654 & -15239 & -15009 \\ 
$\alpha\beta\omega$-GARCH, GICSS, Student-t & 17260 & -34490 & -34391 & 15872 & -31681 & -31477 & -14768 & -29529 & \textbf{-29509} & 7809 & -15548 & -15319 \\ 
$\alpha\beta\omega$-GARCH, GNPCPM, Gaussian & 17140 & -34267 & -34221 & 15713 & -31405 & -31332 & -14693 & -29380 & -29360 & 7564 & -15113 & -15067 \\ 
$\alpha\beta\omega$-GARCH, GNPCPM, Student-t & 17246 & -34478 & \textbf{-34432} & 15838 & -31655 & \textbf{-31582} & -14768 & -29529 & \textbf{-29509} & 7716 & -15419 & \textbf{-15373} \\ 

\hline 
\end{tabular}
}
\caption{The log-likelihood, AIC, and BIC associated with each model on the four stock indexes. The bold text shows the model which gives the best fit as measured by each criterion.}
\label{tab:garchfit}
\end{table}

Having completed our preliminary analysis of the change points, we now fit the change point GARCH models. As discussed in Section \ref{sec:garch}, we consider several different types of models. As a benchmark, we fit GARCH model with no change points, using both the Gaussian and Student t distributions for the error distributions. Next, we fit the $\omega-GARCH$ and $\alpha\beta\omega-GARCH$ where the segment boundaries correspond to the change points found by the ICSS and NPCPM algorithms.

As a final modeling remark, it is possible that the large number of change points found by the ICSS and NPCPM algorithms are an artifact of the two-stage process we are using to fit the models. Both of these change point detection algorithms assume that the observations are independent, however since we are applying these algorithms before the GARCH model is fit, it is possible for the autocorrelation in the volatility to cause an unusually high number of false positives.  We therefore also considered a three-stage model fitting procedure of the following form: first, a GARCH(1,1) model is fit to the return series and the conditional variance $\sigma_t^2$ is estimated on each day. This is then used to standardize the observations via the transformation $y_t = r_t/\sigma_t$. If the GARCH model correctly fits the data and does not contain any change points then these transformed variables should be independent with variance $1$. The ICSS and NPCPM algorithms are then applied to these transformed variables to find any change points. Finally, separate GARCH models are fit within each of the discovered regimes.  When using this procedure, the number of change points found drops substantially with the ICSS procedure finding 3, 7, 0 and 8 change points in the four indexes respectively, with the NPCPM finding 1,2,0 and 1, both of which are substantial reductions compared to the number found when running the algorithm on the raw sequences. In the following discussion we will refer to the models fit in this manner as GICSS and GNPCPM respectively, to denote the fact that the algorithms are applied to the residuals from an initial GARCH fit rather than to the raw data.

\subsection{GARCH Model Fitting}

In order to compare which models best describe the data, we use the standard Akaike Information Criterion (AIC) and Bayesian Information Criterion (BIC) model comparison metrics \cite{Burnham2004, Yang2005}.   Both of these measure how well a model fits the observed data, based on the likelihood of the data under the model, with a penalization for the number of parameters in the model. This penalization is necessary in order to prevent overfitting,  and balance out the increase in the likelihood that an overparameterized model will generally have. The AIC is defined as:

$$AIC = 2k - 2 \log (L)$$

where $L$ is the likelihood of the model under the MLE parameter estimate, and $k$ is the number of parameters in the model. Similarly, the BIC is defined as: 

$$BIC = k \log (n) - 2 \log (L),$$

where $L$ and $k$ are as before, and $n$ is the number of observations. With both measures a $\textbf{low value}$ indicates a better fit. The practical difference between the two criterion is that the BIC penalizes model parameters to a greater degree than the AIC. There is some controversy over which of the two criteria is more appropriate, see \cite{Burnham2004} for a review. We choose to report both, and the results are shown in Table \ref{tab:garchfit}.We can draw several conclusions:

\begin{itemize}
\item The models using Student-t errors distribution consistently outperform the Gaussian models. This suggests that even with the time-varying volatility allowed by the GARCH process, the Gaussian distribution still cannot adequately capture the heavy tailed nature of these series. Similar results have been noted by \cite{He2010}.

\item The best fits in terms of likelihood are given by the various $\alpha\beta\omega-GARCH$ models. This shows the advantage of incorporating structural breaks into the GARCH framework

\item The full  $\omega \alpha \beta-GARCH$ models which allow all parameters to vary generally outperform the more parsimonious $\omega -GARCH$ models, even when factoring in the likelihood penalty imposed by BIC/AIC

\item According to the AIC performance measure, the $\omega \alpha \beta-GARCH$ model using NPCPM change points is the best fitting model for every stock index. Although the ICSS methods give comparable values for the likelihood, the exaggerated number of change-points they produce means that give a poorer fit overall. 

\item When the BIC performance measure is used instead, the three stage model where the NPCPM algorithm is applied the residuals of an initial GARCH fit gives the best BIC for three of the four indexes. The exception is the Nikkei index for which no change points were found by either the ICSS or NPCPM methods when applying these to the GARCH residuals, in which case their BIC results are tied. 
\end{itemize}

In summary, the fact that the NPCPM algorithm does not assume Gaussianity means that it is more robust to outliers than the ICSS, and this results in more parsimonious change detection when using daily returns. This is reflected in the performance criteria used to assess model fit, which shows that it gives improved overall results. We would hence generally recommend the $\omega \alpha \beta-GARCH$ model using the Student-t error distribution in conjunction with the NPCPM algorithm when modeling volatility.

We note that we could also have compared these different methods for volatility estimating by treating them as predictive models and comparing their out-of-sample forecasting errors according to some standard criteria such as the mean squared error (MSE). However the question of whether change point models are appropriate for short-term volatility forecasting is still controversial, and there is some evidence \cite{Covarrubias2006} that standard GARCH models may perform better for this purpose depending on the precise performance measure which is used.  Because our main concern is volatility estimation rather than forecasting, we prefer to avoid this issue and use performance measures which relate only to (penalized) model fit.

\subsection{Further Analysis}
After detecting the change points in the above section, it is potentially interesting to investigate whether they correspond to events in the real world. In this section we examine  the Dow Jones index in more detail. Using the NPCPM GARCH algorith,  there were 12 change points detected. For each of the dates at which the changes occurred, we searched through news headlines from the week immediately before and after to find whether any major events occured which may be related. For $5$ of the change points, we managed to find significant economic events which occured within several days and may have been the cause of the volatility shifts:

\begin{itemize}

\item 26th July 2003: two days earlier on the 24th July, the S\&P credit rating agency cut the rating of California bonds from A to BBB.

\item 19th July 2007: the following week, the Dow Jones index experienced a substantial 2.3\% drop over concerns about the housing and credit markets. The volatility increase may have anticipated this.

\item 15th September 2008: on this date Lehmann Brothers filed for bankruptcy, the event which signaled the start of the recent financial crisis.

\item 2nd June 2009: on the previous day, General Motors filed for bankruptcy.

\item 8th August 2011: three days earlier, S\&P downgraded the credit rating of the United States.

\end{itemize}

\begin{table}
\centering
\begin{tabular}{c c}
\hline
Regime & Volatility \\
\hline
2nd January 1991 - 16th May 1991&0.011\\
17h May 1991 - 30th December 1996&0.007\\
31st December 1996 - 16th June 2002&0.012\\
17th June 2002 - 23rd September 2002&0.020\\
24th September 2002 - 17th October 2002&0.028\\
18th October 2002 - 25th July 2003&0.013\\
26th July 2003 - 16th August 2006&0.007\\
17th August 2006 - 18th July 2007&0.006\\
19th July 2007 - 14th September 2008&0.013\\
15th September 2008 - 9th December 2008&0.042\\
10th December 2008 - 1st June 2009&0.020\\
2nd June 2009 - 7th August 2011&0.010\\
8th August 2011 - 16th November 2011&0.019\\
\hline
\end{tabular}
\caption{Unconditional volatility in each segment found by the NPCPM algorithm}
\label{tab:vols}
\end{table}

For the remaining $7$ change points we did not find any specific events. For reference, these were 17th May 1991, 31st December 1996, 17th June 2002, 24th September 2002, 18th October 2002, 17th August 2006, and 10th December 2008. This seems slightly puzzling since the first two change points in 1991 and 1996 correspond to very clear structural change points which can be seen in Figure \ref{fig:gbp05mood}, with the second one marking a pronounced switch from a period of low volatility to high volatility. Since there are no specific associated news events,  it is possible that these changes occurred in response to longer term trends in the markets or economic system rather than being responses to specific events. For example, the bursting of the internet bubble caused a prolonged stock market downturn during 2002, during which the Dow Jones lost almost 17\% of its value, with most of this occurring between May and October. It seems probable that the change points found by NPCPM in June and September are caused by this, even though are are no high profile news events around these dates specifically.

To investigate further, Table \ref{tab:vols} shows the unconditional volatility in each of the segments. It can be seen that the most volatile period was unsurprisingly the three month period immediately following the bankruptcy of Lehmann Brothers in late 2008. After this, volatility decreased but was still high compared to the historical average. The other sustained period of high volatility occurred during 2002 and lasted from June to October. Since this corresponds quite closely to the stock market downturn, it seems reasonable to assume that it was an underlying factor which may have caused the discovered change points around the start and end of this period.

\section{Conclusions}
Many financial applications require an accurate estimate of the historical volatility of specified financial instruments. For example, certain types of derivatives are priced using the realized volatility, such as the popular Merton model which is used to price Credit Default Swaps and is usually estimated by using the volatility of the stock price as an input variable \cite{Jones1984}. Volatility calculations also feature extensively in risk management, with GARCH models finding regular use within  traditional Value-at-Risk (VaR) analysis \cite{Engle2001}. Similarly, accurate volatility estimation is the first step in computing the correlation between financial instruments, which is a central task in portfolio optimization \cite{Elton2009}.

The ICSS-GARCH algorithm has been widely used to model the time varying volatility commonly found in financial returns. In this methodology, the ICSS algorithm is first used to segment the series based on discovered change points, before a GARCH model is fit to each segment. However, ICSS is very sensitive to heavy tailed data, and can flag for spurious change points when used in this setting. This is unfortunate since heavy-tailed behaviour is typical in financial data, and this has limited the use of the algorithm to the study of weekly returns, where large daily price movements are smoothed out. In order to work with daily data, we have introduced an alternative algorithm where we replace ICSS with a test utilizing ideas from nonparametric statistics. Our experimental analysis shows that this generally gives a better fit to daily data, as measured by several standard model selection techniques.

\bibliographystyle{abbrv}	
\bibliography{jabref}	
\end{document}